\documentclass[prl,twocolumn,showpacs,amsmath,amssymb]{revtex4}

\usepackage{graphicx}
\usepackage{bm}
\usepackage{epsf}
\def\be{\begin{equation}}
\def\ee{\end{equation}}
\def\bea{\begin{eqnarray}}
\def\eea{\end{eqnarray}}

\begin{document}
\title{Tunneling into a Luttinger liquid revisited}
\author{D. N. Aristov$^{1,2}$}
\author{A. P. Dmitriev$^{3,2}$}
\author{I. V. Gornyi$^{2,3}$}
\author{V. Yu. Kachorovskii$^{3,2}$}
\author{D. G. Polyakov$^{2}$}
\author{P. W\"olfle$^{4,2}$}
\affiliation{
$^{1}$Petersburg Nuclear Physics Institute, 188300 St.Petersburg, Russia
\\
$^{2}$Institut f\"ur Nanotechnologie, Karlsruhe Institute of Technology,
76021 Karlsruhe, Germany
\\
$^{3}$A.F.Ioffe Physico-Technical Institute,
194021 St.Petersburg, Russia
\\
$^{4}$Institut f\"ur Theorie der kondensierten Materie, Karlsruhe Institute of Technology, 76128 Karlsruhe, Germany
}
\date{\today}

\begin{abstract}
We study how electron-electron interactions renormalize tunneling into a Luttinger liquid beyond the lowest order of perturbation in the tunneling amplitude. We find that the conventional fixed point has a finite basin of attraction only in the point contact model, but a finite size of the contact makes it generically unstable to the tunneling-induced breakup of the liquid into two independent parts. In the course of renormalization to the nonperturbative-in-tunneling fixed point, the tunneling conductance may show a nonmonotonic behavior with temperature or bias voltage.
\end{abstract}

\maketitle

Electron tunneling into a correlated many-electron system is one of the most essential tools to probe the nature of the correlations. A key concept here is that the tunneling density of states reflects how difficult it is for electronic states to rearrange themselves to accommodate the extra charge of the tunneling electron. The hallmark of strong correlations is the ``zero-bias anomaly" (ZBA) \cite{nazarov09}---the nonlinear behavior of the tunneling current as a function of the bias voltage---resulting from a singularity in the tunneling density of states at the Fermi energy.

The prototype model and the best-understood example of a strongly correlated metallic state is the Luttinger liquid (LL) in one-dimensional electron systems \cite{giamarchi04}. The LL behavior has been observed in nanowires---of which the most prominent examples are the carbon nanotubes and the semiconductor quantum wires---through the power-law suppression \cite{bockrath99,auslaender02} of the tunneling conductance with decreasing bias voltage and/or temperature.

\begin{figure}
\centerline{\includegraphics[width=0.6\columnwidth]{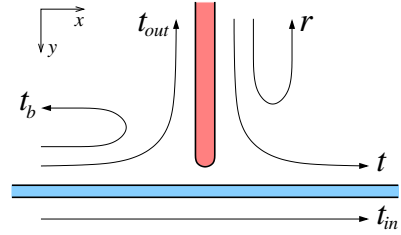}}
\caption{Junction between a quantum wire (horizontal) and a tunnel electrode (vertical). The arrows denote the scattering processes (and their amplitudes) for incoming electrons.}
\label{f1}
\end{figure}

When using the term ``tunneling", we often have at the back of our minds that the tunneling probes the properties of the system into which the electron tunnels ``noninvasively", i.e., the tunneling amplitude is infinitesimally small. A more subtle and complete understanding of the interplay of strong correlations and tunneling emerges when the latter is treated beyond the lowest order of perturbation theory. It is the purpose of this paper to study the junction between a LL and a tunnel electrode (Fig.~\ref{f1}) for arbitrary strength of tunneling. Apart from the conceptual interest, the ``three-way junction" is considered a key element for device engineering in nanoelectronics.

Our main result is the phase diagram for the interaction-induced renormalization of the parameters of the tunnel junction. We show that a homogeneous LL is actually generically {\it unstable} at low energies to arbitrarily weak tunneling and the breakup into two independent semi-infinite wires. The ZBA at the true (stable) fixed point (FP) is strongly enhanced. For sufficiently weak interaction or sufficiently strong tunnel coupling, the tunnel conductance actually {\it grows} with decreasing energy scale (temperature, bias voltage) before it reaches maximum and starts to renormalize toward zero.

Let us specify the model. The Hamiltonian reads $H=H_{\rm w}+H_{\rm e}+H_{\rm tun}$, where $(\hbar=1)$
$H_{\rm w}=\sum_\mu\int\!dx\left(-i\mu v\psi^\dagger_\mu\partial_x\psi_\mu+{1\over 2}V_0n_\mu n_{-\mu}\right)$
describes the LL wire (for compactness, we focus
here on the spinless case), $\mu=\pm$ denotes
electrons moving to the right (+) and to the left ($-$). The electron operator $\psi$, defined in the complete basis of scattering states of the three-way junction, is decomposed inside the wire as a sum of chiral components: $\psi(x)=\sum_\mu\psi_\mu(x)$, and $n_\mu(x)$ is the density fluctuation in the channel $\mu$. We assume that the interaction potential between electrons in the wire is screened by a nearby metallic gate and take it to be point-like with the zero-momentum component $V_0$. The interaction between electrons
with the same $\mu$ is then fully incorporated in
the renormalization of
the velocity $v$ \cite{giamarchi04}.
We focus in this paper on the case of small $\alpha=V_0/2\pi v$. We assume that the interaction is present in a finite region of length $L$ around the tunnel contact, which models a LL wire connected to noninteracting leads. For simplicity, we represent here the tunnel electrode as a noninteracting semi-infinite wire described by $H_{\rm e}=-iv_{\rm e}\sum_\mu\mu\int_{-\infty}^0\!dy\,\psi^\dagger_\mu\partial_y\psi_\mu$, where $\psi_\mu$ are the chiral components of the electron operator $\psi(y)$
on the half-axis of $y$. In the absence of tunneling, the scattering states at $y<0$ are characterized by the phase of the reflection coefficient $r=e^{i\phi_r}$.

The term $H_{\rm tun}$ describes the tunnel junction. We start by considering the simplest and commonly used model (``point contact") for the tunneling Hamiltonian: $H_{\rm tun}=t_0\psi^\dagger (y=-0)\psi(x=0)+{\rm H.c.}$, take $t_0$ to be real and put the phase $\phi_r=0$. In the absence of interaction, the scattering amplitudes (Fig.~\ref{f1}) can be shown to obey $t_b=-\rho$, $t_{in}=1-\rho$, $t_{out}=t=-i\,{\rm sgn}(t_0)[2\rho(1-\rho)]^{1/2}$, $r=1-2\rho$, where $\rho=2|t_0|^2/(vv_{\rm e}+2|t_0|^2)$. We see that the virtual transitions from the wire into the tunnel contact and back lead to backscattering in the wire.

As is commonly known \cite{kane92}, a weak backscattering amplitude in a LL is renormalized by interaction and behaves as $(\Lambda/|\epsilon|)^{1-K}$, where the energy $\epsilon$ is counted from the Fermi level, $\Lambda$ is the ultraviolet cutoff, and the Luttinger parameter $K=(1-\alpha)^{1/2}(1+\alpha)^{-1/2}\simeq 1-\alpha$ for small $\alpha$. At first glance, one might expect that the amplitude $t_b$ is renormalized similarly. However, the fact that unitarity in the tunnel junction is imposed on the $3\times 3$ (and not $2\times 2$) scattering matrix has dramatic consequences for the renormalization. We find (see below) that, to second order in $\alpha$, the exact-in-$\rho$ beta-function $\beta (\rho)={\partial \rho}/{\partial {\cal L}}$, where ${\cal L}=\ln (\Lambda/|\epsilon|)$ for $|\epsilon|\gg v/L$, for the point tunnel contact reads:
\begin{equation}
\beta(\rho)=\alpha \rho(1-\rho)[\,\rho-\alpha (1-2\rho)(1-\rho+\rho^2)/2\,]~.
\label{1a}
\end{equation}
For $\rho\ll 1$ it reduces to
\begin{equation}
\beta(\rho)\simeq -\alpha^2\rho/2+\alpha \rho^2~,\quad \rho\ll 1~.
\label{1}
\end{equation}
It is most important that $\beta (\rho)$ in Eq.~(\ref{1}) does not contain the term $\alpha \rho$ linear in both $\alpha$ and $\rho$, which would give the renormalization of $t_b$ similar to that for the weak impurity in a LL wire. To better understand this, recall that the renormalization can be described \cite{yue94} in terms of scattering off the Friedel oscillations produced by the scatterer. A subtlety of the point tunnel junction is that both $t_b$ and $t_{in}$---in contrast to the impurity---are real and, as a result, the contributions to $\beta (\rho)$ at order ${\cal O}(\alpha \rho)$ from the Friedel oscillations that ``dress" the junction at $x<0$ and $x>0$ exactly cancel each other.

\begin{figure}
\centerline{\includegraphics[width=0.7\columnwidth]{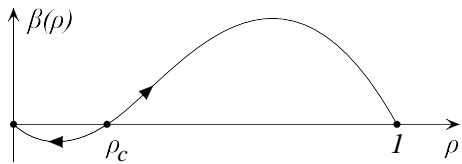}}
\caption{Renormalization group flows (schematic) for the point tunnel junction. Two stable FPs at $\rho=0$ and $\rho=1$ describe the homogeneous  wire and the breakup of the wire into two independent semi-infinite pieces, respectively. Tunneling into the wire is blocked at both points.
}
\label{f2}
\end{figure}

The first term in Eq.~(\ref{1}) means the usual suppression \cite{giamarchi04} of the tunneling density of states $\propto (|\epsilon|/\Lambda)^{(1-K)^2/2K}$ in a homogeneous LL. The second term comes from the renormalization of backscattering along the wire at first order in $\alpha$, as found earlier in Ref.~\cite{lal02}. What seems to have not been discussed in the literature
is that the two terms have opposite signs (for $\alpha>0$), so that $\beta (\rho)$ for small $\alpha$ vanishes at $\rho=\rho_c\simeq\alpha/2$, which signifies an unstable FP (Fig.~\ref{f2}) leading to the phase transition separating two phases with $\rho=0$ and $\rho=1$ \cite{remark2}. For $\rho,\alpha\ll 1$ the solution of Eq.~(\ref{1}) gives
\begin{equation}
\rho\simeq {\rho_0\over 2\rho_0/\alpha + (1-2\rho_0/\alpha)(\Lambda/|\epsilon|)^{\alpha^2/2}}~,
\label{2}
\end{equation}
where $\rho_0$ is the bare value of $\rho$.

We see that the tunneling is suppressed by interaction only if its bare amplitude is sufficiently small (or, equivalently, if the interaction is sufficiently strong). Otherwise, according to Eq.~(\ref{1a}), the tunneling constant $t_0$ monotonically {\it increases} in the course of renormalization.
The tunneling transparency $G_{\rm t}=2|t|^2$, however, shows a nonmonotonic behavior with increasing $t_0$:
\begin{equation}
G_{\rm t}=4\rho(1-\rho)=8vv_{\rm e}|t_0|^2/(vv_{\rm e}+2|t_0|^2)^2~.
\label{2a}
\end{equation}
If $\rho_c<1/2$, $G_{\rm t}$ grows until it reaches maximum $G_{\rm t}=1$ (which means a completely transparent contact with $r=0$) at $\rho=1/2$ and decreases at larger $\rho$, eventually vanishing at $\rho=1$. By contrast, $G_{\rm t}$ for $\rho<\rho_c$ decreases monotonically in the course of renormalization.

\begin{figure}
\centerline{\includegraphics[width=0.7\columnwidth]{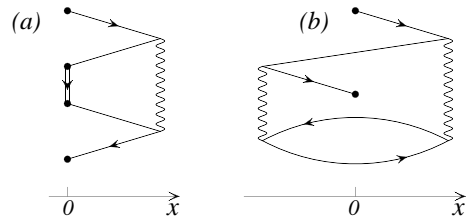}}
\caption{Lowest-order scattering processes leading to the renormalization of the reflection amplitude $r$. Dots: tunneling amplitude $t_0$. Wavy lines: bare interaction $\alpha$ between the right and left movers. The straight single and double lines denote the electron Green function in the wire and in the tunnel electrode, respectively. }
\label{f3}
\end{figure}

It is instructive to analyze the scattering processes diagrammatically in the energy-space representation; in particular, this makes it easy to count powers of $t_0$ and $\alpha$. The first and second terms in Eq.~(\ref{1}) correspond, respectively, to diagrams (b) and (a) in Fig.~\ref{f3} for the amplitude $r$. In Fig.~\ref{f3}(a), tunneling with the amplitude $t_0$ into the right-moving state is followed by scattering at first order in $\alpha$ off the Friedel oscillation, whose amplitude is proportional to $t_b\sim {\cal O}(t_0^2)$. Returning to the tunnel electrode costs one more power of $t_0$. Altogether, this gives $\alpha |t_0|^4$, which explains the origin of the second term in Eq.~(\ref{1}) (and---since the process of first order in $\alpha$ necessarily includes backscattering at the contact---it also explains why the term of order $\alpha \rho$ is absent). The other scattering process [Fig.~\ref{f3}(b)] is of second order in $\alpha$ because of the creation of an electron-hole pair, but requires only two powers of $t_0$ (only to get into the wire and come back): this gives $\alpha^2|t_0|^2$.

\begin{figure}
\centerline{\includegraphics[width=0.85\columnwidth]{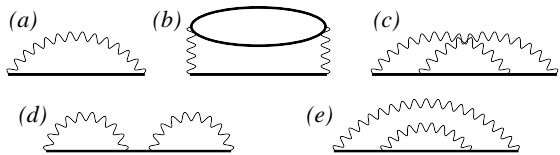}}
\caption{Skeleton diagrams (a) and (b) contribute to the exact-in-$\rho$ $\beta$-function at second order in $\alpha$, diagrams (c)-(e) do not. The thick straight lines denote the noninteracting electron Green function fully dressed by the tunneling vertices. }
\label{f4}
\end{figure}

More complicated diagrams of higher orders in $\rho$ are constructed similarly: to second order in $\alpha$ they are compactly represented in Fig.~\ref{f4}, where the thick lines denote the noninteracting Green function dressed in all possible ways by the tunnel vertices. Importantly, only one-loop diagrams (a) and (b) in Fig.~\ref{f4} contribute to the $\beta$-function to second order in $\alpha$. Diagrams (c)-(e) cancel all singular interaction-induced terms in the $S$-matrix except those given by the renormalization group equation.
In particular, the perturbative corrections to $r$ which come from Figs.~4(a) and (b) read
$\delta r_{(4a)}\to -\alpha tt_{out}t_b^*{\cal L}$ and $\delta r_{(4b)}\to -\alpha^2 tt_{out}(t_b^*|t_b|^2+t_{in}^*|t_{in}|^2){\cal L}/2$. Calculating diagrams (a) and (b) for other amplitudes gives similar expressions, which---when written for the point tunnel contact in terms of $\rho$---all reduce to the single equation, Eq.~(1).

As follows from Eq.~(\ref{1a}), tunneling into the wire is blocked at both FPs $\rho=0$ and $\rho=1$: $G_{\rm t}$ vanishes as $|\epsilon|^{\alpha^2/2}$ at $\rho\to 0$ and as $|\epsilon|^\alpha$ at $\rho\to 1$. We see that the ZBA for small $\alpha$ is strongly enhanced in the latter case. The difference in the exponents reflects the difference in the state of the wire: although the tunnel contact is decoupled from the wire at both FPs, the wire at $\rho=0$ is homogeneous, while at $\rho=1$ it is broken up into two disconnected pieces
(the exponent coincides then with that for tunneling into the end of a semi-infinite wire \cite{giamarchi04}).

\begin{figure}
\vspace{-7mm}
\centerline{\includegraphics[width=0.75\columnwidth]{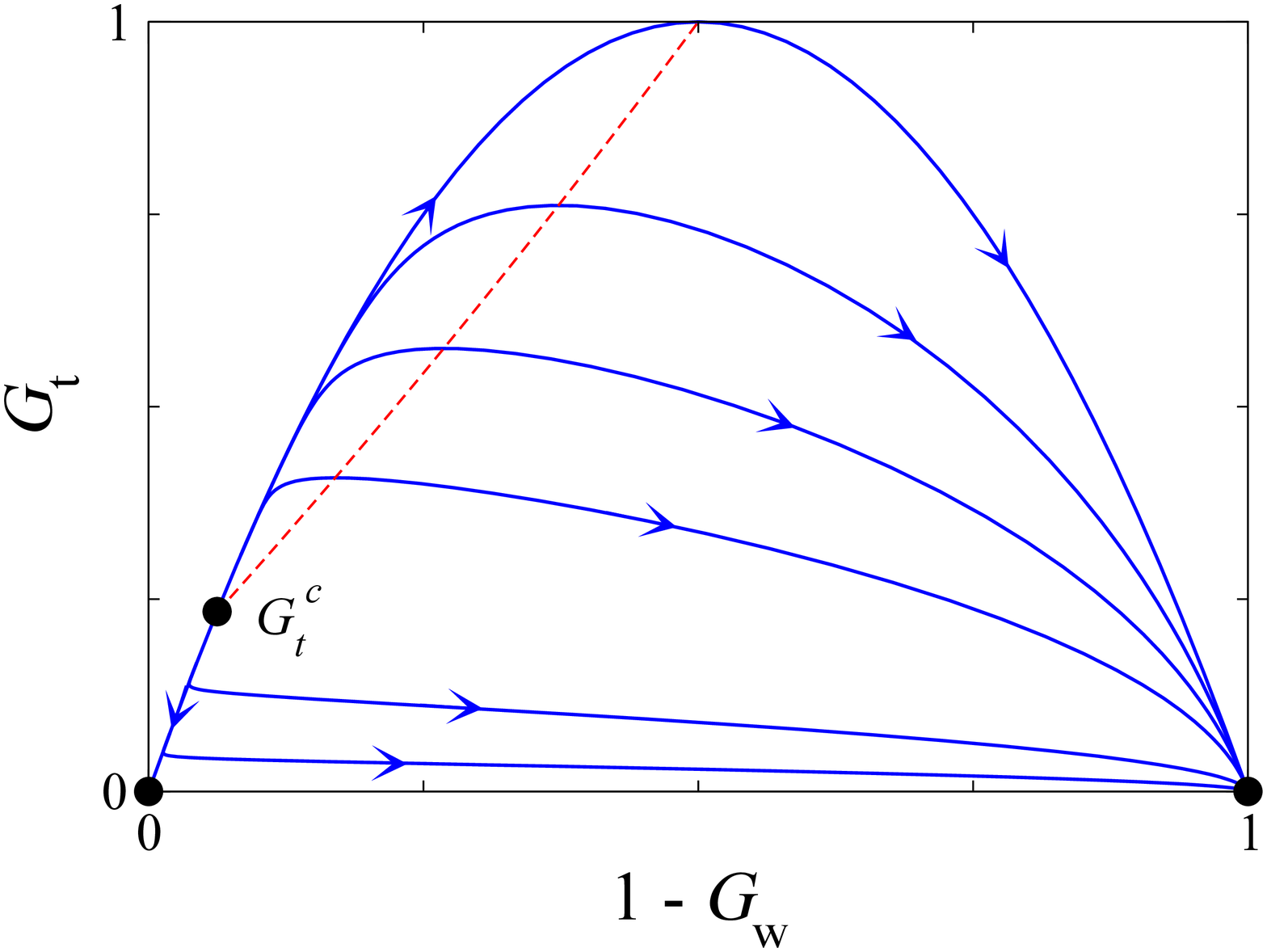}}
\caption{Renormalization group flows for the tunnel junction in the $G_{\rm t}$--$G_{\rm w}$ plane for $\alpha=0.15$. The only stable FP is at $G_{\rm t}=G_{\rm w}=0$ (breakup of the junction into three parts). The unstable FP at $G_{\rm t}=G_{\rm t}^c$ corresponds to the point $\rho_c$ in Fig.~\ref{f2}. The dashed line separates the regions in which $G_{\rm t}$ grows (above) and falls off (below) with decreasing energy.}
\label{f5}
\end{figure}

We now turn to a more general form of the tunnel coupling and show that the FP at $\rho=0$, obtained above for the commonly used model of the point contact, is actually generically unstable. The general parametrization of the $3\times 3$ $S$-matrix obeying time reversal symmetry and mirror symmetry $x\leftrightarrow -x$ (in particular, at an extended tunnel contact) gives for the moduli (which we focus on here) of the scattering amplitudes:
\begin{equation}
|t_b|^2=\rho^2~,\quad|t|^2=|t_{out}|^2=2\rho(\cos\varphi -\rho) ~,
\label{3}
\end{equation}
where $\rho$ and $\varphi$ are constrained by the condition $0<\rho<\cos\varphi<1$ and the unitarity condition reads $2|t|^2+|r|^2=|t_{out}|^2+|t_{in}|^2+|t_b|^2=1$. The point contact with real $t_0$ corresponds to $\varphi=0$.
The wire decoupled from the tunnel electrode corresponds to $\rho=\cos\varphi$.

Diagrams (a) and (b) in Fig.~\ref{f4}, where the thick lines denote now the ``dressing" by the tunnel vertices described by Eqs.~(\ref{3}), yield a set of two {\it closed} \cite{remark3} flow equations for arbitrary $\rho$ and $\varphi$. In the limit of small $\rho\ll\alpha,\cos\varphi$, the $\beta$-functions read
\cite{safi09}:
$\partial \rho/\partial {\cal L}\simeq\alpha \rho(\sin^2\varphi-\alpha/2)$ and $\partial \varphi/\partial {\cal L}\simeq (\alpha/2) \sin 2\varphi$. It follows that the FP $(\rho=0,\varphi=0)$ is actually a saddle point in $(\rho,\varphi)$ space. The key difference from Eq.~(\ref{1a}) is that the function  $\partial \rho/\partial {\cal L}$ for small $\rho$ now acquires the term $\alpha \rho\sin^2\varphi$.
This means that at $\varphi\neq 0$ the cancellation of scattering on the Friedel oscillations at order ${\cal O}(\alpha \rho)$ [cf.\ the discussion below Eq.~(\ref{1})] is no longer exact. Written in terms of the dimensionless conductances
$G_{\rm t}=2|t|^2$ and $G_{\rm w}=|t_{in}|^2+|t|^2/2$ \cite{remark4}, the flow equations are:
\begin{equation}
\begin{aligned}
\partial G_{\rm t}/\partial{\cal L}&=\alpha G_{\rm t}(f_1 + \alpha f_2)~, \\
\partial G_{\rm w}/\partial{\cal L}&=\alpha (f_3 + \alpha f_4)~,
\end{aligned}
\label{4}
\end{equation}
where $f_1=-1+G_{\rm t}/2+G_{\rm w}$, $f_2=-1/2+G_{\rm t}(3-G_{\rm t})/8+G_{\rm w}(1-G_{\rm w})$, $f_3=G_{\rm t}(1+G_{\rm w})/4-2G_{\rm w}(1-G_{\rm w})$, $f_4=(1-2G_{\rm w})[G_{\rm t}(4+G_{\rm t})/32-G_{\rm w}(1-G_{\rm w})]$. The flow stops at $\epsilon$ given by $T$, bias, or $v/L$, whichever is larger.

\begin{figure}
\vspace{-9mm}
\centerline{
\includegraphics[width=0.9\columnwidth]{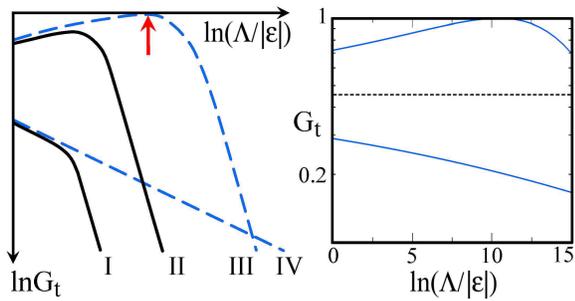}}
\vspace{-7mm}
\caption{
{\it Left panel:} schematics of the renormalization of the tunnel conductance $G_{\rm t}$ on the log-log scale for the generic tunnel contact with $\varphi\neq 0$ (solid lines) and the point contact (dashed). Curves I and IV (II and III) correspond to the bare conductance $G_{{\rm t}0}$ below (above) the dashed line in Fig.~\ref{f5}. The arrow indicates the point at which the contact is reflectionless. The flow of $G_{\rm t}$ stops at the largest scale of $\epsilon$ among $T$, $v/L$, and the voltage between the tunnel electrode and the wire. {\it Right panel:} the tunnel conductance of the point contact for $\alpha=0.4$ and $G_{{\rm t}0}$ above (upper curve) and just below (lower) the critical conductance $G_{\rm t}^c\simeq 0.45$ (dashed line). The criticality retards the development of the ZBA. }
\label{f6}
\end{figure}

The flow diagram for $G_{\rm t} $ and $G_{\rm w}$ is shown in Fig.~\ref{f5}.
Two limiting curves are the line $G_{\rm t}=4G_{\rm w}(1-G_{\rm w})$ which describes the point tunnel junction, with the unstable FP from Fig.~\ref{f2} at $G_{\rm t}=G_{\rm t}^c\simeq 2\alpha$, and the line $G_{\rm t}=0$ which describes a decoupled tunnel electrode. The FP $(G_{\rm t}=0,G_{\rm w}=1)$ is seen to be unstable to the decrease of $G_{\rm w}$. It is this point that describes the commonly expected outcome of the renormalization: a homogeneous LL with tunneling blocked by the ZBA. In fact, this point has a finite basin of attraction (the line $0<G_{\rm t}<G_{\rm t}^c$) only in the model of the point tunnel contact (which, albeit being widely used, is not generic in this sense). Generically, all flows are toward the point $G_{\rm t}=G_{\rm w}=0$ which describes the breakup of the junction into {\it three} disconnected parts.

For $1-G_{\rm w}\ll 1$ and the bare values $G_{{\rm t}0},1-G_{{\rm w}0}\ll \alpha$, Eqs.~(\ref{4}) can be linearized, which gives:
\begin{equation}
1-G_{\rm w}-G_{\rm t}/4=\left(G_{{\rm t}0}/G_{\rm t}\right)^\kappa\left(1-G_{{\rm w}0}-G_{{\rm t}0}/ 4\right)~,
\label{5}
\end{equation}
where $\kappa\simeq 2(2-\alpha)/\alpha$. We see that if the flow is initially close to the limiting curve $1-G_{\rm w}\simeq G_{\rm t}/4$ predicted by the point contact model, it shows eventually a kink, after which the ``deviation" from the point contact model begins to grow sharply ($\kappa\sim\alpha^{-1}$ for small $\alpha$), see Fig.~\ref{f5}. The critical exponents change at the kink: before it both $G_{\rm t}$ and $1-G_{\rm w}$ decrease as $|\epsilon|^{\alpha^2/2}$ \cite{aristov_unp}, after it $G_{\rm t}$ continues to follow this power law but $1-G_{\rm w}$ increases as $|\epsilon|^{-2\alpha}$. Near the FP at $G_{\rm w}=0$, Eqs.~(\ref{4}) predict that the exponent of $G_{\rm t}$ also changes and both $G_{\rm t}$ and $G_{\rm w}$ vanish at the FP as $|\epsilon|^\alpha$ \cite{remark5}. The scaling of $G_{\rm t,w}\propto |\epsilon|^\alpha$ describes tunneling into the end of a semi-infinite LL wire \cite{aristov_unp}, which means enhancement (curve I in Fig.~\ref{f6}) of the ZBA as compared to the conventional picture of tunneling into a homogeneous LL (curve IV). If the bare values $G_{{\rm t}0},1-G_{{\rm w}0}$ lie above the dashed line in Fig.~\ref{f5}, interactions first make the tunnel contact more transparent (curves II and III in Fig.~\ref{f6}), so that both $G_{\rm t}$ and $1-G_{\rm w}$ first grow, but eventually the flow is attracted to the same FP $G_{\rm t}=G_{\rm w}=0$. Note that the development of the ZBA can be strongly hindered in the vicinity of $G_{\rm t}^c$, as shown in Fig.~\ref{f6} (right).

To summarize, we have shown that the picture of tunneling into a LL is qualitatively modified when the tunneling amplitude is not treated as infinitesimally small. The conventional FP has a finite basin of attraction only in the model of the point tunnel contact, but taking a finite size of the contact (or any perturbation induced by the contact in the wire) into account makes it unstable. Generically, at the only stable FP the junction breaks up into three disconnected parts. Flowing toward this FP, the tunnel conductance may behave nonmonotonically with bias voltage or temperature. Our predictions can be verified by systematically varying the distance to the tunnel electrode in experiments on carbon nanotubes or semiconductor nanowires.

We thank S.~Das, Y.~Oreg, I.~Safi, and O.~Yevtushenko for interesting discussions. The work was supported by the DFG/CFN, the EuroHORCs/ESF, the RAS, GIF Grant No.\ 965, the RFBR, and the DFG-RFBR.


\vspace{-4mm}

\begin{thebibliography}{99}

\vspace{-4mm}

\bibitem{nazarov09} Yu.V.~Nazarov and Ya.M.~Blanter, {\it Quantum Transport: Introduction to Nanoscience} (Cambridge University Press, Cambridge, 2009).

\bibitem{giamarchi04} T.~Giamarchi, {\it Quantum Physics in One Dimension} (Oxford University Press, Oxford, 2004).

\bibitem{bockrath99} M.~Bockrath {\it et al.}, Nature {\bf 397}, 598 (1999); Z.~Yao {\it et al.}, {\it ibid.} {\bf 402}, 273
(1999).

\bibitem{auslaender02} O.M.~Auslaender {\it et al.}, Science {\bf 295}, 825 (2002); E.~Levy {\it et al.}, Phys.\ Rev.\ Lett.\ {\bf 97}, 196802 (2006); Y.~Jompol {\it et al.}, Science {\bf 325}, 597 (2009).

\bibitem{kane92} C.L.~Kane and M.P.A.~Fisher, Phys.\ Rev.\ B {\bf 46}, 15233 (1992).

\bibitem{yue94} D.~Yue, L.I.~Glazman, and K.A.~Matveev, Phys.\ Rev.\ B {\bf 49}, 1966 (1994).

\bibitem{lal02} S.~Lal, S.~Rao, and D.~Sen, Phys.\ Rev.\ B {\bf 66}, 165327 (2002); S.~Das, S.~Rao, and D.~Sen, {\it ibid.} {\bf 70}, 085318 (2004).

\bibitem{remark2} In Ref.~\cite{lal02}, this FP was missed because the calculation of the $\beta$-function was restricted to the first order in $\alpha$. The FP at $\rho\neq 0,1$ obtained in Ref.~\cite{lal02} was {\it entirely} due to the presence of interaction in the LL tunnel electrode. The latter FP was also described in X.~Barnab{\'e}-Th{\'e}riault {\it et al.}, Phys.\ Rev.\ B {\bf 71}, 205327 (2005)---whose method is restricted to the same level of accuracy---for the case of tunneling between identical LLs. For a discussion of the junction of identical LLs and the stable (breakup) FP at $\rho=1$, see also R.~Egger {\it et al.}, New J. Phys.\ {\bf 5}, 117 (2003); M.~Oshikawa, C.~Chamon, and I.~Affleck, J. Stat.\ Mech.\ (2006) P02008; A.~Agarwal {\it et al.}, Phys.\ Rev.\ Lett.\ {\bf 103}, 026401 (2009).

\bibitem{remark3} The evolution of the phases of the $S$-matrix is completely determined by the flow of the moduli of its elements.

\bibitem{safi09} In the limit of $\rho\to 0$, the linearized-in-$\rho$ flow equations were also proposed in I.~Safi, arXiv:0906.2363 [see Eqs.~(127) there]---in a form which depends crucially on an unspecified ``nonuniversal" coefficient $c$ (in contrast, all coefficients in our calculation are well-defined and give $c=\alpha/2$). Note that the linearized-in-$\rho$ equations yield the growth of $\rho$ for $\varphi\neq 0$ but this does not guarantee the flow to the breakup FP at $\rho=1$. To obtain the breakup, starting from the vicinity of the FP at $\rho=0$, one should go beyond the linear-in-$\rho$ approximation [see Eqs.~(6)].

\bibitem{remark4} The ``wire conductance" $G_{\rm w}$ gives the current in a biased wire under the condition that the applied potentials are such that no current flows through the tunnel electrode.

\bibitem{aristov_unp} Following the method of D.N.~Aristov and P.~W{\"o}lfle, Phys.\ Rev.\ B {\bf 80}, 045109 (2009), the one-loop contributions can be summed up to infinite order in $\alpha$ to give a phase portrait similar to that in Fig.~\ref{f5} and the critical exponents of $G_{\rm t,w}$ equal to $(1-K)^2/2K$ and $(1-K)/K$ for the FPs at $G_{\rm w}=1$ and $G_{\rm w}=0$, respectively (D.N.~Aristov and P.~W{\"o}lfle, unpublished).

\bibitem{remark5} Note that the wire transparency $|t_{in}|^2$ vanishes at this FP with the doubled exponent as $|\epsilon|^{2\alpha}$.

\end{thebibliography}
\end{document}